\documentclass[12pt,a4paper]{article}
\usepackage[T1]{fontenc} 
\usepackage[dvipdfmx]{graphicx}
\usepackage{bm,caption,latexsym,amssymb,amsfonts,mathtools,here}
\usepackage{bm,latexsym,amsfonts,mathtools}
\usepackage{amssymb, amsmath, comment}
\usepackage{color}
\usepackage{hyperref}
\usepackage{booktabs}
\usepackage{mathrsfs}
\usepackage{wrapfig}
\usepackage{setspace}
\numberwithin{equation}{section}
\renewcommand{\baselinestretch}{1.5}

\usepackage[sorting=none,doi=false]{biblatex}

\bibliography{reference}

\begin{document}%%%%%%%%%%%%%%%%%%%%%%%%%%%%%%%%%%%%%%%%%%%
\begin{titlepage}
\unitlength = 1mm
\begin{flushright}
KOBE-COSMO-22-13
\end{flushright}

\vskip 1cm
\begin{center}

{ \large \textbf{Anisotropic warm inflation }}

\vspace{1.8cm}
Sugumi Kanno$^*$, Ann Mukuno$^\flat$, Jiro Soda$^{\flat}$, and Kazushige Ueda$^*$

\vspace{1cm}

\shortstack[l]
{\it $^*$ Department of Physics, Kyushu University, Fukuoka 819-0395, Japan \\ 
\it $^\flat$ Department of Physics, Kobe University, Kobe 657-8501, Japan}

\vskip 4.0cm

{\large Abstract}\\
\end{center}

Anisotropic inflation is a model succeeded in explaining statistical anisotropy. 
Warm inflation is a model succeeded in providing a mechanism of reheating during inflation.
We study anisotropic warm inflation focusing on the cosmic no-hair conjecture. 
In the anisotropic warm inflation, the condition for making anisotropy survived is clarified.
By assuming a constant value for the
dissipation ratio, we find exact solutions of power-law anisotropic warm inflation, and investigate the phase space structure of general solutions.  
It turns out that whether the anisotropy during inflation survives or not depends on the competition of the potential that drive anisotropic inflation against dissipation of an inflaton field. Anisotropic warm inflation will be realized if the decaying process is not efficient.
%strength} of decaying process from inflaton field into matter fields. \textcolor{red}{Anisotropic warm inflation will be realized if the decaying process is weak enough.}
%If the process is very slow, the anisotropic warm inflation is realized. If the process is slow, the anisotropy tends to be hard to grow enough in the duration of inflation. If the process is rapid, no anisotropic inflation occurs.

\vspace{1.0cm}
\end{titlepage}

%%%%%%%%%%%%%%%%%%%%%%%%%%%%%%%%%%%%%
%%%%%%%%%%%%%%%%%%%%%%%%%%%%%%%%%%%%%
\hrule height 0.075mm depth 0.075mm width 165mm
\tableofcontents
\vspace{1.0cm}
\hrule height 0.075mm depth 0.075mm width 165mm
%%%%%%%%%%%%%%%%%%%%%%%%%%%%%%%%%%%%%
%%%%%%%%%%%%%%%%%%%%%%%%%%%%%%%%%%%%%
\section{Introduction}%%%%%%%%%%%%%%%%%%%%%%%%%%%%%%%%%%%
Cosmological inflation driven by a single scalar field, the inflaton, has succeeded in accounting for the current cosmological observations, such as the cosmic microwave background (CMB) radiation~\cite{Planck:2013jfk}. When the inflaton field rolls down the inflaton potential very slowly compared to the expansion of the universe, inflation occurs. Especially in the slow-roll limit, the inflaton potential energy is regarded as a  cosmological constant.
In the presence of the cosmological constant, there is a  cosmic no-hair theorem for homogeneous and anisotropic spacetimes~\cite{Wald:1983ky}. 
The cosmic no-hair theorem states that if a cosmological spacetime obeys the Einstein equation with the positive cosmological constant, (1) the energy density of ordinary matter vanishes, (2)  anisotropy of the spacetime vanishes, and (3) spatial curvature vanishes (except for the Bianchi IX spacetimes with a very large curvature scale compared to the Hubble scale).
It had been believed that the above statements hold even away from the slow-roll limit and for inhomogeneous spacetimes, and it is called the cosmic no-hair conjecture. If the cosmic no-hair conjecture is true, the conjecture guarantees that the power spectra of primordial fluctuations are statistically homogeneous, isotropic and scale free~\cite{Soda:2012zm}.

In the era of high-precision observation, 
however, we need to take into account the deviation from the cosmological constant. Indeed, the interaction of the inflaton with other matter fields destabilize the inflation of a single scalar field and qualitatively different inflation models emerge. 
For instance, particle production during inflation may lead to a counterexample to the statement (1), which is known as warm inflation~\cite{Berera:1995ie,Berera:1995wh,Berera:2008ar}. 
In the warm inflation, the decay from the inflaton to the matter leads to dissipation. This induces a thermal bath described by a radiation energy density during inflation. 
This radiation energy density never vanishes during inflation. Soon after the proposal of the warm inflation, however, it has been pointed out that quantum corrections at finite temperature tend to destroy the warm inflation scenario~\cite{Yokoyama:1998ju}. Then, several models of warm inflation were constructed to evade the problem by introducing symmetries such as the supersymmetry~\cite{Berera:1998px}~\cite{Berera:2003kg}~\cite{Bastero:2016qru}~\cite{Bastero:2019gao}. Recently,
a minimal warm inflation model is also proposed in the context of axion inflation~\cite{Berghaus:2019whh}. In the model, the shift symmetry of the axion is utilized to prohibit the quantum corrections. 
On the other hand, a counterexample to the statement (2) is also found as a model of anisotropic inflation by considering a dynamical cosmological constant due to a gauge field coupled with the slow-rolling scalar field~\cite{Watanabe:2009ct}. In this model, statistical anisotropy can be produced during inflation~\cite{Dulaney:2010sq,Gumrukcuoglu:2010yc,Watanabe:2010fh}.
The mechanism of destabilization of the isotropic universe 
in the model of anisotropic inflation has been studied in detail~\cite{Chen:2021nkf,Chen:2022ccf}.
The anisotropic inflation has been investigated in various  models (see review articles \cite{Soda:2012zm,Maleknejad:2012fw}).
As to the statement (3), it would be difficult to 
construct a counterexample without a fine tuning.
Hence, we will not discuss this case in this paper.

From the perspective of the cosmic no-hair conjecture, it is interesting to consider the models which violate the statement (1) and (2) at the same time. 
A primary question is whether anisotropic warm inflation occurs or not.  
%In the usual slow-roll inflation, 
For instance, in the case of chaotic inflation, the solution of the slow-roll inflation becomes an attractor solution in the phase space of the dynamical system of the inflaton field. 

However, we show that the additional matter fields of the radiation energy density and gauge fields give rise to thermal and non-thermal destabilization of the conventional slow-roll inflation, respectively. 
Hence, it is interesting to clarify the phase space  structure of the dynamical system in the presence of the radiation energy density and the gauge fields. 
We show that a fixed point that corresponds to an attractor solution of the anisotropic warm inflation exists.
Depending on model parameters, 
the property of the fixed point changes. 
Indeed, in the presence of dissipation of the inflaton field, warm inflation becomes attractor in the absence of gauge fields. And the attractor changes to a saddle point once the gauge field emerges for an appropriate coupling between the inflaton and the gauge fields. Then the anisotropy appears after the warm inflation if the inflation lasts forever.
However, for a wide range of model parameters, inflation ends before the anisotropic inflation starts.
If the dissipation is strong enough compared to the effect of the gauge fields, the anisotropy disappears. 
Thus, we find that whether the anisotropy during inflation survives or not depends on the strength of the dissipation. All calculations and estimates are
based on the assumption of considering a constant value for the
dissipation ratio. In fact,  warm inflation models have instead a dissipation
ratio that is not constant but has dependencies on the temperature
and, in some cases, also on the inflaton amplitude. However, in the slow-roll limit, the dissipation ratio is almost constant during inflation. Hence, assumption that dissipation ratio is constant is legitimate to clarify the condition for the existence of a solution of anisotropic warm inflation.  To support the validity of the assumption, we show  qualitative feature does not change even for time dependent dissipation ratio in the Appendix.

The paper is organized as follows. In section 2, we review warm inflation briefly. In particular, we clarify slow-roll conditions in the context of warm inflation. In section 3, we consider warm inflation in the presence of a gauge filed. We examine the condition for making anisotropy survived in the warm inflation.
In section 4, we consider exactly solvable power-law inflation models in order to clarify the phase space structure of the models. In section 5, we perform a dynamical system approach. We analytically obtain some fixed points in the dynamical system. We numerically study dynamical flow and reveal the role of the dissipation in the anisotropic inflation.

\section{A review of warm inflation}
\label{section2}
%%%%%%%%%%%%%%%%%%%%%%%%%%%%%%%%%%%%%%%%%%%
%A benefit of warm inflation is to provide a mechanism of reheating during inflation. The point is that a small coupling of the inflaton field to other light fields gives rise to particle production and generate a thermal bath.
%Recently, Berghaus et al. presented a simple model of warm inflation which avoids the usual problem of thermal backreaction on the inflaton potential by considering an axion-like coupling to gauge bosons~\cite{Berghaus:2019whh}. 
Warm inflation has several attractive features. 
The thermal friction can alleviate the required flatness of the potential.
It also provides a mechanism of reheating during inflation.
Moreover, since dissipation can modify the growth of inflaton fluctuations,
imprints on the CMB fluctuations can be used to probe the interactions 
between the inflaton and other particles. Here, we give a brief review of the warm inflation.

We begin with Einstein Hilbert action coupled to an inflaton field $\phi$ and a matter field $\psi_{\rm matter}$ described by the free Lagrangian ${\cal L}_{\rm  free}$. The interaction between the inflaton and the matter fields is represented by  ${\cal L}_{\rm  int}$ such as
\begin{eqnarray}
S=\int d^4 x \sqrt{-g}
\left[
\frac{M_{\rm pl}^2}{2} R -\frac{1}{2}(\partial_\mu \phi)(\partial^\mu \phi) -V(\phi) 
+ {\cal L}_{\rm  free}\left( \psi_{\rm matter} \right)
+ {\cal L}_{\rm  int}\left(\phi , \psi_{\rm matter} \right)
\right],
\label{action1}
\end{eqnarray}
where $g$ is the determinant of the metric $g_{\mu\nu}$,  $M_{\rm pl}$ is the Planck constant $M_{\rm pl}^{-2}=8\pi G$.

We will take the metric to be homogeneous and isotropic, that is,
\begin{eqnarray}
ds^2=-dt^2+a(t)^2\,\delta_{ij}\,dx^idx^j\,,
\end{eqnarray}
where $a(t)$ is the scale factor.
The dissipation of the inflaton field due to the coupling to the matter field leads to particle production. This induces a thermal bath sourcing effective friction during inflation.~\cite{Hosoya:1983ke}. The effect of the dissipation comes in the Klein-Gordon equation in the form of effective friction such as
\begin{eqnarray}
\ddot{\phi}+3H\left(1+Q\right)\dot{\phi}+V'\left(\phi\right)=0 \ ,
\label{eom1}
\end{eqnarray}
where $H\equiv\dot{a}/a$ is the Hubble parameter and a dot denotes derivative with respect to time $t$. The  dissipation of the inflaton field is characterized by  $\Upsilon=3HQ$ where $Q$ is dimensionless parameter and we call it dissipation
ratio in the following.
In general, the dissipation depends on the inflaton field, temperature, and the mass of the inflaton fields. The resulting  thermal radiation bath is represented by radiation energy density $\rho_{\rm R}$. Then the Friedman equation becomes
\begin{eqnarray}
H^2=\frac{1}{3M_{\rm pl}^2} \left(V(\phi)+\frac{1}{2}\dot{\phi}^2+\rho_{\rm R}\right) \ .
\label{eom2}
\end{eqnarray}
The conservation of energy-momentum as a consequence of the  contracted Bianchi identity yields
\begin{eqnarray}
\dot{\rho}_{\rm R}+4H\rho_{\rm R}=\Upsilon \dot{\phi}^2 \ .
\label{eom3}
\end{eqnarray}
We see that radiation energy density is continuously 
sourced from the dissipation of the inflaton field.

Let us consider slow-roll inflation. In the slow-roll approximation, we impose the conditions 
$|\ddot{\phi}|\ll H\,|\dot{\phi}|$ and $\dot{\phi}^2\ll V(\phi)$ in Eqs.~(\ref{eom1}) and (\ref{eom2}). In the presence of the dissipation of inflaton field, we assume the conditions $\rho_{\rm R} \ll V(\phi)$ and
$\dot{\rho}_{\rm R} \ll H\rho_{\rm R}$ in Eqs.~(\ref{eom2}) and (\ref{eom3}) hold.
Then, Eqs. (\ref{eom1}), (\ref{eom2}), and (\ref{eom3}) are simplified to
\begin{eqnarray}
&&3H(1+Q)\,\dot{\phi}+V'(\phi)=0 \, ,
\label{EOM1}\\
&&H^2=\frac{V(\phi)}{3M_{\rm pl}^2} \, ,
\label{EOM2}\\
&&4H\rho_{\rm R}=\Upsilon \dot{\phi}^2\, ,
\label{EOM3}
\end{eqnarray}
where a prime denotes a derivative with respect to $\phi$.
As standard slow-roll inflation is a dynamical attractor, we expect that the above slow-roll inflation with dissipation is also a dynamical attractor. For a self-consistent approximation, we need to impose the following slow-roll conditions
\begin{eqnarray}
&&\epsilon_V\equiv \frac{M_{\rm pl}^2}{2(1+Q)}
\left(
\frac{V'(\phi)}{V(\phi)}
\right)^2 \ll 1,
\label{ev}\\
&&\eta_V\equiv\frac{M_{\rm pl}^2}{1+Q}\frac{V''(\phi)}{V(\phi)}\ll 1,\\
&&\beta\equiv\frac{M_{\rm pl}^2}{1+Q}\frac{Q'\,V'(\phi)}{Q\,V(\phi)}\ll 1 ,
\label{beta}
\end{eqnarray}
where we assumed no temperature dependence on $Q$ in the slow-roll conditions for simplicity. In the following, we also assume that $Q$ has no $\phi$-dependence in Eq.~(\ref{beta}) because $Q^\prime$ has to be zero in the limit of $\beta\rightarrow0$.

In general, thermal backreaction on the inflaton potential violate the above slow-roll conditions~\cite{Yokoyama:1998ju}. 
%\textcolor{red}{However, (many other models of warm inflation have been constructed that do not have thermal backreaction problems to prohibit the quantum corrections by introducƒing symmetries. )}
%\textcolor{blue}{warm inflation models evading the thermal backreaction problems have been constructed by introducing symmetries to prohibit quantum thermal corrections.
%For example, supersymmetry is utilized in ~\cite{Berera:1998px} and ~\cite{Berera:2003kg}. Warm little inflation exploits a similar mechanism as little Higgs scenarios ~\cite{Bastero:2016qru,Bastero:2019gao}. The shift-symmetry of axion is used in minimal warm inflation~ \cite{Berghaus:2019whh}.}
However, such thermal backreaction can be avoided by introducing symmetries. For example, \cite{Berera:1998px} and \cite{Berera:2003kg} use supersymmetry, \cite{Bastero:2016qru,Bastero:2019gao} propose warm little inflation by introducing little Higgs scenario with similar symmetries, and \cite{Berghaus:2019whh} makes use of shift-symmetry of axion in minimal warm inflation.

As seen in Eqs.~(\ref{EOM1}), (\ref{EOM2}) and (\ref{EOM3}), the system reduces to standard slow-roll inflation in the absence of the dissipation ($\Upsilon=0$). In the presence of the dissipation, even if the inflation started without radiation energy density ($\rho_{\rm R}=0$), Eq.~(\ref{eom3}) tells us that $\rho_{\rm R}$ rapidly increases and reach the attractor value $\sim Q\dot{\phi}^2$. 
Interestingly, in this case, the radiation energy density survives during inflation and the amount survived is estimated as $\rho_{\rm R} \sim \epsilon_V V(\phi)$ for $Q>1$ where Eqs.~(\ref{EOM1}), (\ref{EOM2}), (\ref{EOM3}) and (\ref{ev}) are used. 
That is, this is a counterexample to the cosmic no-hair conjecture that the energy density of ordinary matter vanishes during inflation.
In terms of temperature $T$, Stefan-Boltzmann law gives $\rho_{\rm R} \sim T^4$.  By using Eq.~(\ref{EOM2}), we find $\rho_R\sim\epsilon_V V(\phi)=\epsilon_V M_{\rm pl}^2 H^2  $. Then we have $T\sim \epsilon_V^{1/4} \sqrt{M_{\rm pl}/H} H$. 
Thus,  for low energy inflation ($H<M_{\rm pl}$), warm inflation $T>H$ is realized. 

The number of e-folds is calculated as 
\begin{eqnarray}
N\sim \frac{1}{M_{\rm pl}^2} \int_{\phi_f}^{\phi_i} d\phi
\left(1+Q\right) \frac{V}{V'} \,,
\end{eqnarray}
where $\phi_i$ and $\phi_f$ are the initial and the final value of the inflaton field. This tells us that the existence of dissipation enables inflation to occur even in the steep potential  ($M_{\rm pl}V^\prime\gg V$).

\section{Anisotropic warm inflation}
%%%%%%%%%%%%%%%%%%%%%%%%%%%%%%%%%%%%%%%%%%%
As a counterexample to cosmic no-hair conjecture, anisotropic inflation is realized by considering a dynamical cosmological constant due to a gauge field coupled with a slow-rolling scalar field~\cite{Watanabe:2009ct}. In this section, we consider warm inflation in the context of~\cite{Watanabe:2009ct} and see if the anisotropy can  survive or not. 
Different models of anisotropic inflation have been studied in the context of warm inflation in the literature~\cite{Sharif:2013zwa,Sharif:2014aaa}.
However, those models are known to be unstable~\cite{Himmetoglu:2008zp} or anisotropy disappears~\cite{Maleknejad:2011jr}.

On top of the terms ${\cal L}_{\rm  free}(\psi_{\rm matter})$ and ${\cal L}_{\rm  int}(\phi,\psi_{\rm matter})$, we introduce a gauge field $A_\mu$ to the action Eq.~(\ref{action1}) in the form
\begin{eqnarray}
S=\int d^4 x \sqrt{-g}
\left[
\frac{M_{\rm pl}^2}{2} R -\frac{1}{2}(\partial_\mu \phi)(\partial^\mu \phi) -V(\phi) -\frac{1}{4}f^2(\phi)F_{\mu\nu}F^{\mu\nu}
\right],
\end{eqnarray}
where the field strength of the gauge field is defined by $F_{\mu\nu}=\partial_\mu A_\nu-\partial_\nu A_\mu$ and  $f(\phi)$ is the gauge kinetic function representing the coupling between the inflaton field and the U(1) gauge field. The U(1) gauge field neither thermalize nor contribute to the dissipation $\Upsilon$. Instead, the U(1) gauge field plays the role of producing anisotropic expansion.
For an appropriate gauge kinetic function, it was shown that the energy density of the gauge fields survives  during inflation against the cosmic no-hair conjecture in~\cite{Watanabe:2009ct}. Thanks to the gauge invariance, we can choose the gauge $A_0=0$.
Without loss of generality, we
can take the $x$-axis in the direction of the gauge field  $A_\mu=(0,A(t),0,0)$ and $\phi=\phi(t)$. For simplicity, we assume that the direction of the gauge field does not change in time. In order to be consistent with these  setups, we consider anisotropic metric in the form
\begin{eqnarray}
ds^2=-dt^2+e^{2\alpha(t)}
\left[
e^{-4\sigma(t)}dx^2+e^{2\sigma(t)}  (dy^2+dz^2)
\right],
\label{metric}
\end{eqnarray}
where $\sigma$ represents a deviation from isotropy.
As shown in \cite{Soda:2012zm} and \cite{Watanabe:2009ct}, anisotropic inflation occurs if $f(\phi)$ and $V(\phi)$ satisfy
\begin{eqnarray}
 \frac{M_{\rm pl}^2}{2}\frac{f'}{f} \frac{V'}{V}\geq 1 \ .
\end{eqnarray}
Here the equality gives
\begin{eqnarray}
   f(\phi) = \exp \left[ \frac{2}{M_{\rm pl}^2}\int  \frac{V}{V'}d\phi \right] \ .
\end{eqnarray}
 And the inequality is satisfied if parameter $c>1$ is introduced such as
%It is convenient to take the following form of the gauge kinetic function 
\begin{eqnarray}
f(\phi)=\exp\left[{\frac{2c}{M_{\rm pl}^2}\int \frac{V}{V'}d\phi}\right]\,,
\label{f}
\end{eqnarray}
%\textcolor{red}{Here, the inequality is replaced by the condition $c>1$. In this study, we used this form of the gauge kinetic function as a representative one.}
%by introducing a constant parameter $c$. 
Then 
Maxwell's equations are 
\begin{eqnarray}
  \frac{d}{dt} \left(f^2(\phi)\, 
  e^{\alpha + 4\sigma }\dot{A}
  \right)
  =0 \ ,
\end{eqnarray}
and the solution is obtained by
\begin{eqnarray}
  \dot{A} 
  = p_Af^{-2}(\phi)\, 
  e^{-\alpha -4\sigma}\ ,
  \label{A}
\end{eqnarray}
where $p_A$ is a constant of integration.
Substituting Eq.~(\ref{A}) into Klein-Gordon equation, Einstein's equations and the conservation of energy-momentum, we obtain the following basic equations
\begin{eqnarray}
&&\ddot{\phi}+3\dot{\alpha}(1+Q)\dot{\phi}+V'(\phi)-p_A^2 f^{-3}(\phi)f'(\phi)e^{-4\alpha-4\sigma}=0, 
\label{kg}\\
&&\dot{\alpha}^2=\dot{\sigma}^2+\frac{1}{3M_{\rm pl}^2} \left[
\frac{1}{2}\dot{\phi}^2+V(\phi)+\frac{p_A^2}{2}f^{-2}(\phi)e^{-4\alpha-4\sigma}+\rho_{\rm R}
\right],
\label{friedmann}\\
&&\ddot{\sigma}=-3\dot{\alpha}\dot{\sigma}+\frac{p_A^2}{3M_{\rm pl}^2}f^{-2}(\phi)e^{-4\alpha-4\sigma} \ ,
\label{anisotropy}\\
&&\ddot{\alpha}=-3\dot{\alpha}^2+\frac{V(\phi)}{M_{\rm pl}^2}+\frac{p_A^2}{6M_{\rm pl}^2}f^{-2}(\phi)e^{-4\alpha-4\sigma}+\frac{\rho_{\rm R}}{3\rm M_{\rm pl}^2} \ ,\\
&&\dot\rho_{\rm R}+4\dot\alpha\rho_{\rm R}=3\dot\alpha Q\dot\phi^2
\label{radiation}\ .
\end{eqnarray}
By virtue of the Bianchi identity, one of the above equations becomes redundant. Here the total energy density consists of $\rho_\phi$, $\rho_A$ and $\rho_{\rm R}$. The $\rho_\phi$ and $\rho_A$ are expressed as
\begin{eqnarray}
\rho_\phi&=&\frac{1}{2}\dot{\phi}^2+V(\phi)\,,\\
\rho_A&=&\frac{p_A^2}{2}f^{-2}(\phi)\,e^{-4\alpha-4\sigma}=\frac{p_A^2}{2}e^{-4\frac{c}{M_{\rm pl}^2}\int \frac{V}{V^\prime}\,d\phi\,-4\alpha-4\sigma} \,.
\end{eqnarray}
Now we consider slow-roll inflation. Besides the slow-roll approximation performed in the isotropic inflation in Section.~\ref{section2}, in the anisotropic inflation, we assume the inflaton potential energy $V(\phi)$ is dominant in the total energy density. The expansion rate of anisotropy $\dot{\sigma}$ is supposed to be small compared to the Hubble parameter $\dot{\alpha}$. 
Then, Eqs.(\ref{kg}), (\ref{friedmann}), (\ref{anisotropy}) and (\ref{radiation}) are simplified as
\begin{eqnarray}
&&3\dot{\alpha}(1+Q)\dot{\phi}+V'(\phi)
-p_A^2 f^{-3}(\phi)f'(\phi)e^{-4\alpha-4\sigma}=0\ ,
\label{AKG}
\\
&&\dot{\alpha}^2=\frac{1}{3M_{\rm pl}^2}V(\phi) \ ,\label{AFeq}\\
&&3\dot{\alpha}\dot{\sigma}=\frac{p_A^2}{3M_{\rm pl}^2}f^{-2}(\phi)\,e^{-4\alpha-4\sigma}\ ,
\label{Aniso}
\\
&& 4\dot{\alpha}\rho_{\rm R}=3\dot\alpha Q\dot{\phi}^2 \ .
\end{eqnarray}
As shown below, we find $f^{-2}(\phi)\propto e^{4\alpha}$ in the r.h.s of Eq.~(\ref{Aniso}). Hence, 
the $\dot{\sigma}$ does not decay even if $\alpha$ increases. 
Combining Eqs.(\ref{AKG}) with (\ref{AFeq}), we have
\begin{eqnarray}
(1+Q)\frac{d\phi}{d\alpha} 
+M_{\rm pl}^2 \frac{V'}{V}
-p_A^2 \frac{2c}{V'}\, e^{-4\frac{c}{M_{\rm pl}^2}\int \frac{V}{V^\prime}\,d\phi\,-4\alpha-4\sigma}   =0 \ ,
\end{eqnarray}
where we used Eq.~(\ref{f}). The solution of the above differential equation is
\begin{eqnarray}
e^{-4\frac{c}{M_{\rm pl}^2}\int \frac{V}{V^\prime}\,d\phi\,-4\alpha-4\sigma}
=\frac{M_{\rm pl}^2\left(c-1-Q\right)}{2c^2p_A^2}\frac{V^{\prime 2}}{V}
\left[
1+D\,e^{-4\frac{c-1-Q}{1+Q} \alpha}
\right]^{-1}\ ,
\label{rhoA}
\end{eqnarray}
where $D$ is a constant of integration. Note that $Q$ is constant. If the parameter $c$ satisfies the inequality 
\begin{eqnarray}
 c>1+Q \label{ani-condition}\ ,
\end{eqnarray}
the term proportional to $D$ decays as  $\alpha\rightarrow\infty$ and the r.h.s of Eq.~(\ref{rhoA}) becomes a positive constant in the slow-roll limit and expressed as
\begin{eqnarray}
e^{-4\frac{c}{M_{\rm pl}^2}\int \frac{V}{V^\prime}\,d\phi\,-4\alpha-4\sigma}
=\frac{M_{\rm pl}^2\left(c-1-Q\right)}{2c^2p_A^2}\frac{V^{\prime 2}}{V} \ .
\label{relation}
\end{eqnarray}
In this way, $f^{-2}(\phi)=\exp[-\frac{4c}{M^2_{\rm pl}}\int\frac{V}{V^\prime}\,d\phi]\propto e^{4\alpha}$ is shown.

By using Eqs.(\ref{AFeq}) and (\ref{Aniso}), the degree of anisotropy is expressed by 
\begin{eqnarray}
  \frac{\Sigma}{H}
  \equiv \frac{\dot{\sigma}}{\dot{\alpha}}
  = \frac{p_A^2}{3V(\phi)}\,e^{-4\frac{c}{M_{\rm pl}^2}\int \frac{V}{V^\prime}\,d\phi-4\alpha-4\sigma}
  =\frac{2}{3}\frac{\rho_A}{V(\phi)} \ .
\end{eqnarray}
Thus, for $\alpha\rightarrow\infty$, the degree of the anisotropy converges to
\begin{eqnarray}
\frac{\Sigma}{H}
=\frac{1}{3}\frac{(c-1-Q)(1+Q)}{c^2}\epsilon_V 
\ ,
\label{condition}
\end{eqnarray}
where we used Eqs.~(\ref{relation}) and (\ref{ev}). If there is no dissipation $Q=0$, the anisotropy survives during inflation for $c>1$. However, in the presence of dissipation, the condition for making anisotropy survived becomes tight $c>1+Q$.

\section{Power-law anisotropic warm inflation}%%%%%%%%%%%%%%%%%%%%%%%%%%%%%%%%%%%%%%%%%%%
\label{section4}
In this section, we introduce the exponential potential and gauge kinetic function in order to figure out what happens to the dynamics of anisotropic warm inflation using the solvable models. We present some exact solutions without making the slow-roll approximation for the system of Eqs.~(\ref{kg})-(\ref{radiation}) by considering exponential forms of $V(\phi)$ and $f(\phi)$. In this case, inflation never ends.
Without the dissipation, exact solutions of power-law inflation are found in~\cite{Kanno:2010nr,Do:2011zza,Yamamoto:2012tq,Ohashi:2013pca,Ito:2015sxj,Do:2017qyd}.

We consider the potential
\begin{eqnarray}
V(\phi)=V_0\,\exp\left[
\lambda\frac{\phi}{M_{\rm pl}}\right]\,,
\label{potential}
\end{eqnarray}
and the gauge kinetic function
\begin{eqnarray}
f(\phi)=f_0\,\exp\left[\kappa\frac{\phi}{M_{\rm pl}}\right] \ ,
\label{function}
\end{eqnarray}
where $V_0$, $\lambda$, $f_0$ and $\kappa$ are constants.
The metric and the gauge field is given in Eqs.~(\ref{metric}) and (\ref{A}), respectively. 
The system of  equations of motion can be written as
\begin{eqnarray}
&&\ddot{\phi}+3\dot{\alpha}(1+Q)\dot{\phi}
+\frac{\lambda}{M_{\rm pl}}V_0 e^{\lambda\frac{\phi}{M_{\rm pl}}}
-\frac{\kappa}{M_{\rm pl}}p_A^2 f_0^{-2} e^{-2\kappa\frac{\phi}{M_{\rm pl}}}e^{-4\alpha-4\sigma}=0, 
\label{kg2}\\
&&\dot{\alpha}^2=\dot{\sigma}^2+\frac{1}{3M_{\rm pl}^2} \left[
\frac{1}{2}\dot{\phi}^2+V_0 e^{\lambda\frac{\phi}{M_{\rm pl}}}
+\frac{p_A^2}{2}f_0^{-2} e^{-2\kappa\frac{\phi}{M_{\rm pl}}}e^{-4\alpha-4\sigma}+\rho_{\rm R}
\right],
\label{friedmann2}\\
&&\ddot{\sigma}=-3\dot{\alpha}\dot{\sigma}+\frac{p_A^2}{3M_{\rm pl}^2}f_0^{-2} e^{-2\kappa\frac{\phi}{M_{\rm pl}}}e^{-4\alpha-4\sigma},
\label{anisotropy2}\\
&&\ddot{\alpha}=-3\dot{\alpha}^2
+\frac{1}{M_{\rm pl}^2}V_0 e^{\lambda\frac{\phi}{M_{\rm pl}}}+\frac{p_A^2}{6M_{\rm pl}^2}
f_0^{-2} e^{-2\kappa\frac{\phi}{M_{\rm pl}}}e^{-4\alpha-4\sigma}+\frac{\rho_{\rm R}}{3\rm M_{\rm pl}^2},
\label{alpha2}\\
&&\dot\rho_{\rm R}+4\dot\alpha\rho_{\rm R}=3\dot\alpha Q\dot\phi^2.
\label{radiation2}
\end{eqnarray}
The Bianchi identity makes one of the above equations redundant.
By making an ansatz of the form
\begin{eqnarray}
\alpha=\zeta\log M_{\rm pl}\,t,\qquad\sigma=\eta\log M_{\rm pl}\,t,\qquad \phi=M_{\rm pl}\,\omega\log M_{\rm pl}\,t+\phi_0,\qquad \rho_{\rm R}=\frac{B M_{\rm pl}^2 }{t^2} \ ,
\label{ansatz}
\end{eqnarray}
where $\zeta$, $\eta$, $\omega$ and $B$ are dimensionless parameters.
Substituting the above ansatz into Eq.~(\ref{friedmann2}), the following scaling solutions are found to be necessary
\begin{eqnarray}
&&\lambda\omega=-2 \label{Relation-1} \ , \\
&&\kappa\omega+2\eta+2\zeta=1 \ , \label{Relation-2}
\end{eqnarray}
and Eqs.~(\ref{kg2})-(\ref{radiation2}) give
\begin{eqnarray}
&&\omega-3\zeta(1+Q)\omega +\kappa v-\lambda u=0
\label{Aeq-1}\ ,\\
&&-\zeta^2+\eta^2+\frac{1}{6}\omega^2+\frac{1}{3}u+\frac{1}{6}v+\frac{B}{3}=0
\label{Aeq-2} \ ,\\
&&\eta-3\zeta\eta+\frac{1}{3}v=0
\label{Aeq-3} \ ,\\
&&\zeta-3\zeta^2+u+\frac{1}{6}v+\frac{B}{3}=0
\label{Aeq-4} \ ,\\
&&-2B+4\zeta B=3\zeta Q\omega^2  
\label{Aeq-5}\ ,
\end{eqnarray}
where we defined
\begin{eqnarray}
u=\frac{V_0}{M_{\rm pl}^4}
\exp\left[\lambda\frac{\phi_0}{M_{\rm pl}}\right],\qquad v=\frac{p_A^2}
{M_{\rm pl}^4}f_0^{-2}\exp\left[-2\kappa\frac{\phi_0}{M_{\rm pl}}\right] \ .
\end{eqnarray}

In the case of the solution of isotropic inflation, we discard Eqs.(\ref{Relation-2})
and (\ref{Aeq-3}) and set $\eta=0$ and $v=0$. 
Then we find the solution of the form
\begin{eqnarray}
  \omega&=&-\frac{2}{\lambda} \ , \\
  \zeta &=& \frac{\lambda^2 +4(1+Q) +\sqrt{(\lambda^2 +4+4Q)^2 -16 \lambda^2 }}{4\lambda^2} \ ,
  \label{zeta}\\
  B&=& \frac{3}{8\lambda^2}
  \left(\lambda^2+4Q-4+\sqrt{(\lambda^2 +4+4Q)^2 -16 \lambda^2 }\right) \ ,
\end{eqnarray}
where we chose the plus sign in front of square root in Eq.~(\ref{zeta}) so that inflation occurs ($\alpha>0$).
This is the exact solution for power-law warm inflation.

In the case of anisotropic inflation with dissipation, the solution is obtained as follows.
Eqs.(\ref{Relation-1}) and (\ref{Relation-2})
give
\begin{eqnarray}
   \omega= -\frac{2}{\lambda}\ ,\qquad
 \eta=\frac{1}{2}-\zeta+\frac{\kappa}{\lambda}  
 \label{eta}\ .
\end{eqnarray}
Plugging Eq.~(\ref{eta}) into Eq.(\ref{Aeq-3}), we get
\begin{eqnarray}
 v=3  \left(3\zeta -1\right)
 \left(\frac{1}{2}-\zeta+\frac{\kappa}{\lambda}\right) \ .
 \label{v}
\end{eqnarray}
By using Eqs.~(\ref{eta}) and (\ref{v}), Eq.(\ref{Aeq-1}) is found to be
\begin{eqnarray}
  u= -\frac{2}{\lambda^2}
  +6\frac{1+Q}{\lambda^2}\zeta
  +3\frac{\kappa}{\lambda}\left(3\zeta -1\right)
  \left(\frac{1}{2}-\zeta+\frac{\kappa}{\lambda}\right) \ .
  \label{u}
\end{eqnarray}
Eq.(\ref{Aeq-5}) gives
\begin{eqnarray}
    B= \frac{6 Q}{\lambda^2}
    \frac{\zeta}{2\zeta-1}  \ .
    \label{B}
\end{eqnarray}
Substituting Eqs.~(\ref{eta}), (\ref{v}), (\ref{u}) and (\ref{B}) into Eq.(\ref{Aeq-4}), we obtain the equation for $\zeta$ as
\begin{eqnarray}
    \left(3\zeta -1\right)\left[
    \frac{2}{\lambda^2}
    +\frac{4Q}{\lambda^2}\frac{\zeta}{2\zeta-1}
    -3\left(\frac{ 1}{2}+\frac{\kappa}{\lambda}\right)\zeta
    +\left(\frac{1}{2}+3\frac{\kappa}{\lambda} \right)
    \left(\frac{1}{2}+\frac{\kappa}{\lambda}\right)\right]=0 \ .
\end{eqnarray}
The relevant solution is 
\begin{eqnarray}
 \zeta =\frac{4Q+4+2\lambda^2+7\lambda\kappa+6\kappa^2
 +\sqrt{D}}{6\lambda(\lambda+2\kappa)} 
 \label{solution-zeta} \ ,
\end{eqnarray}
where we defined
\begin{eqnarray}
D= 16Q^2+8Q(4+2\lambda^2+7\lambda\kappa+6\kappa^2)
+(4-\lambda^2+\lambda\kappa+6\kappa^2)^2
\ .
\end{eqnarray}
Note that the $V(\phi)$ becomes negative and the system becomes unstable for the minus sign in front of the square root in Eq.~(\ref{solution-zeta}) so we chose plus sign.
When $Q=0$, $D$ becomes
\begin{eqnarray}
  D= (4-\lambda^2+\lambda\kappa+6\kappa^2)^2 \ ,
\end{eqnarray}
which reproduce the result of~\cite{Kanno:2010nr}.
Substituting the solution (\ref{solution-zeta}) into (\ref{eta}), we
obtain
\begin{eqnarray}
  \eta  =
  \frac{-4Q-4+\lambda^2+5\lambda\kappa+6\kappa^2
 -\sqrt{D}}{6\lambda(\lambda+2\kappa)} \ . 
\end{eqnarray}
The explicit form of $u$ is complicated and contain no
new information.
On the other hand, $v$ is given by
\begin{eqnarray}
 v=  \frac{(4+4Q+3\lambda\kappa+6\kappa^2
 +\sqrt{D})(-4-4Q+\lambda^2+5\lambda\kappa+6\kappa^2
 -\sqrt{D})}{4\lambda^2(\lambda+2\kappa)^2} \ .
\end{eqnarray}
The positivity of $v$ gives
the condition for the existence of a solution of anisotropy
\begin{eqnarray}
  -4-4Q+\lambda^2+5\lambda\kappa+6\kappa^2
 -\sqrt{D} >0 \ .
\label{Criterion}
\end{eqnarray}
Note that $Q$ is constant. %Note that
If $Q$ is large enough, this inequality is violated, and the anisotropic inflation never occurs.
The degree of anisotropy is given by
\begin{eqnarray}
 \frac{\Sigma}{H}=\frac{-4Q-4+\lambda^2+5\lambda\kappa+6\kappa^2
 -\sqrt{D}}{4Q+4+2\lambda^2+7\lambda\kappa+6\kappa^2
 +\sqrt{D}}>0 \ .
\end{eqnarray}
Thus, the exact solutions of isotropic and anisotropic inflation in the context of warm inflation are obtained.

\section{Dynamical system approach}
\label{section5}
In the previous section, we obtained particular solutions in the form (\ref{ansatz}).
In order to see the features of general
solutions, we resort to the dynamical system analysis in this section\footnote{The paper~\cite{Rocco:2022} utilized a dynamical system approach in the context of a double scalar field of Warm inflation.}.

To analyze the dynamical feature of the system, we introduce the following dimensionless variables~\cite{Kanno:2010nr,Ito:2015sxj},
\begin{eqnarray}
X=\frac{\dot{\sigma}}{\dot{\alpha}}\,,\qquad
Y=\frac{1}{M_{\rm pl}} \frac{\dot{\phi}}{\dot{\alpha}}\,,\qquad
Z=p_A \frac{f^{-1}(\phi)}{M_{\rm pl}\,\dot{\alpha}} e^{-2\alpha-2\sigma}\,,\qquad
W=\frac{\rho_{\rm R}}{M_{\rm pl}^2 \dot{\alpha}^2} \,,
\end{eqnarray}
Note that the above variables become constant for the particular solutions defined by Eq.~(\ref{ansatz}).
Then, the Hamiltonian constraint equation (\ref{friedmann2}) is written as
\begin{eqnarray}
-\frac{1}{M_{\rm pl}}\frac{V}{\dot{\alpha}^2}
=3(X^2-1)+\frac{1}{2}Y^2+\frac{1}{2} Z^2 +W,
\end{eqnarray}
which gives the condition for the positivity of the potential $V(\phi)$
\begin{eqnarray}
3(X^2-1)+\frac{1}{2}Y^2+\frac{1}{2} Z^2 +W<0 \ .
\end{eqnarray}
This inequality eliminates unphysical solutions that appears below.
In the following, we use the potential $V(\phi)$ and the gauge kinetic function $f(\phi)$ in the form of Eqs.~(\ref{potential}) and (\ref{function}).
Then, the remaining Eqs.~(\ref{kg2}), (\ref{anisotropy2}), (\ref{alpha2}) and (\ref{radiation2}) are rewritten as
\begin{eqnarray}
&&\frac{d X}{d\alpha}=\frac{1}{3}Z^2(X+1)+X\bigl[3(X^2-1)+\frac{1}{2}Y^2+\frac{2}{3}W],
\label{autonomous1}\\
&&\frac{d Y}{d\alpha}=\kappa Z^2+ \lambda\Big[3(X^2-1)+\frac{1}{2}Y^2 +\frac{1}{2}Z^2+W\Big]\nonumber\\
&&\qquad\quad+Y\Big[-3(1+Q)+3X^2+\frac{1}{2}Y^2+\frac{1}{3}Z^2+\frac{2}{3}W \Big] \ ,
\label{autonomous2}\\
&&\frac{d Z}{d\alpha}=Z\bigl[-\kappa Y-2(X+1)+3X^2+\frac{1}{2}Y^2+\frac{1}{3}Z^2+\frac{2}{3}W] 
\ ,
\label{autonomous3}\\
&&\frac{d W}{d\alpha}=3QY^2-4W+6WX^2+WY^2+\frac{2}{3}WZ^2+\frac{4}{3}W^2 \ .
\label{autonomous4}
\end{eqnarray}
The above Eqs.~(\ref{autonomous1}),\,(\ref{autonomous2}),\,(\ref{autonomous3}) and (\ref{autonomous4}) define
an orbit in the four-dimensional phase space $\left(X(\alpha),Y(\alpha),Z(\alpha),W(\alpha) \right)$ once the initial conditions are given. Since l.h.s of these equations becomes zero for the particular solutions discussed in Eq.~(\ref{ansatz}), they correspond to fixed points in this system.
The trajectories never intersect each other except at the fixed points. Therefore, the linear analysis around the fixed point determines the phase space structure.
Depending on the parameters, the number of the fixed points and the properties of the fixed points
change. 

\begin{figure}[H]
\centering
 \includegraphics[width=8cm]{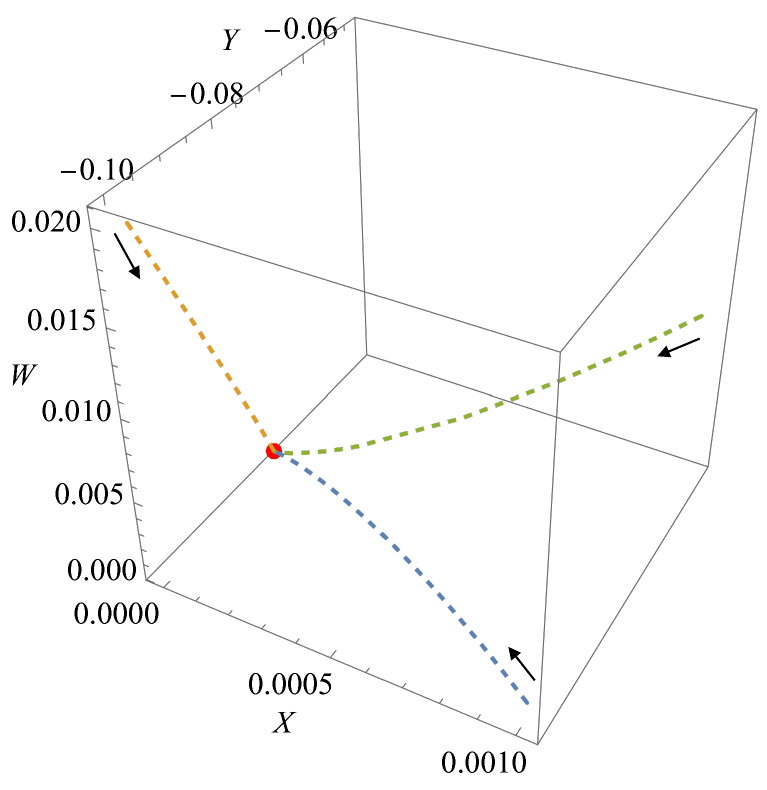}
 \renewcommand{\baselinestretch}{3}
 \caption{The 3D phase space plot in $(X,Y,W)$-space in the absence of the gauge field, i.e. $Z=0$. Trajectories with different initial conditions are depicted for $Q=0.3$.  The initial conditions of blue, yellow and green orbits are set as $(X,Y,W)=(0.001,\,-0.1,\,0.001),~(0,\,-0.1,\,0.02),~(0.001,\,-0.05,\,0.01)$, respectively. The red point at  $(X,Y,W)=(0,\,-0.077,\,0.0013)$ shows the isotropic fixed point of the warm inflation. }
 \label{FigAdd}
\end{figure}

Let us consider the cases that the condition for positivity of $v$ (\ref{Criterion}) is violated. In these cases, there is no anisotropic fixed point. Eq.~(\ref{autonomous4}) tells us that, even if the initial state is vacuum $W=0$, $W$ grows with time because $Q\neq 0$. 
Eventually, the trajectory goes to the isotropic fixed point
\begin{eqnarray}
X&=&0 \, \\
Y&=&-\frac{4(1+Q)+\lambda^2-\sqrt{(4+4Q+\lambda^2)^2-16\lambda^2}}{2\lambda} \ , \\
  Z^2&=&0 \ , \\
  W&=&-\frac{12(1+Q)^2+3\lambda^2(Q-1)-3(1+Q)\sqrt{(4+4Q+\lambda^2)^2-16\lambda^2}}{2\lambda^2} \ .
\end{eqnarray}
In the case of no dissipation $Q=0$, this corresponds to a solution of isotropic power-law inflation $Y=-\lambda,\,W=0$.
In the absence of gauge fields $Z=0$, even if $X\neq 0$ initially, the anisotropy disappears eventually and 
the solution of isotropic warm inflation actually behaves as an attractor. To show this, we plotted the trajectories in the phase space $(X,Y,W)$ in the case of $Z=0$ in FIG \ref{FigAdd}.
However, in the presence of gauge fields, and if the condition (\ref{Criterion}) is satisfied, an anisotropic fixed point exists. In this case, the isotropic fixed point becomes unstable. 
Then, the trajectory goes to
the fixed point
\begin{eqnarray}
  X&=&    \frac{-4Q-4+\lambda^2+5\lambda\kappa+6\kappa^2
 -\sqrt{D}}{4Q+4+2\lambda^2+7\lambda\kappa+6\kappa^2
 +\sqrt{D}}  \ ,  \\  
  Y&=&   -\frac{12(\lambda+2\kappa)}{4Q+4+2\lambda^2+7\lambda\kappa+6\kappa^2
 +\sqrt{D}}  \ ,  \\
  Z^2 &=&  \frac{9(4+4Q+3\lambda\kappa+6\kappa^2
 +\sqrt{D})(-4-4Q+\lambda^2+5\lambda\kappa+6\kappa^2
 -\sqrt{D})}{(4Q+4+2\lambda^2+7\lambda\kappa+6\kappa^2
 +\sqrt{D})^2 }   \ ,\\
  W&=&
    \frac{108 \  Q(\lambda+2\rho)^2}{(4Q+4+2\lambda^2
    +7\lambda\kappa+6\kappa^2
 +\sqrt{D})(4Q+4-\lambda^2+\lambda\kappa+6\kappa^2
 +\sqrt{D})} \ .
\end{eqnarray}
This fixed point corresponds to anisotropic warm inflation obtained in the previous section. 
For $Q=0$, we can reproduce the results in~\cite{Kanno:2010nr}.

\begin{figure}[H]
\centering
 \includegraphics[width=16cm]{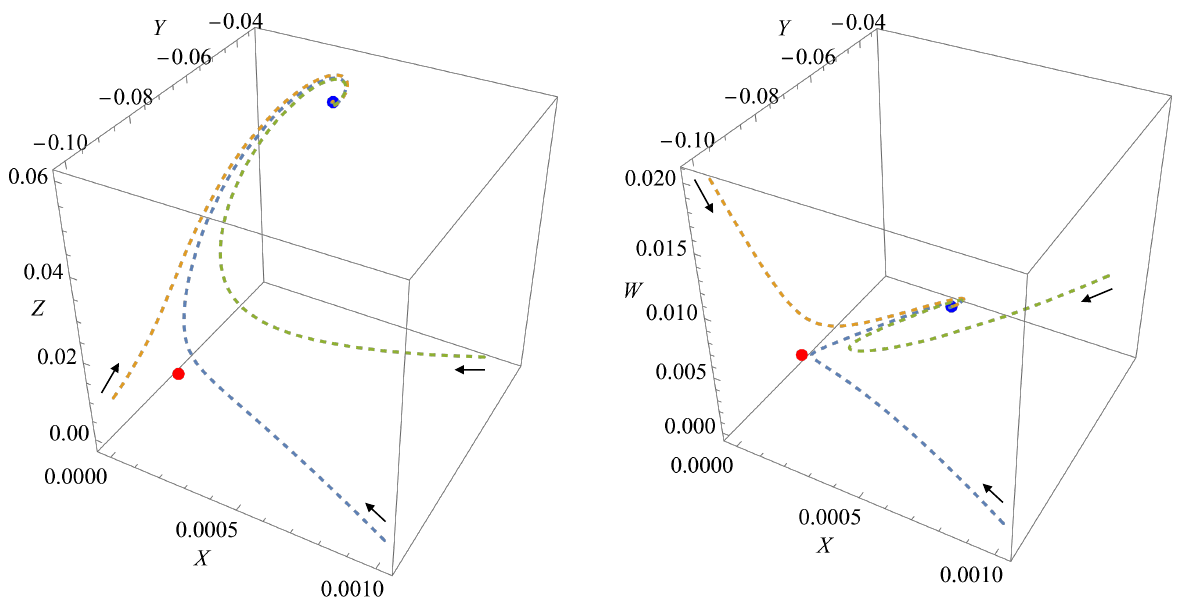}
 \renewcommand{\baselinestretch}{3}
 \caption{3D Phase space plots of attractors in the $(X,Y,Z)$-spaces (Left panel) and $(X,Y,W)$-spaces (Right panel). Trajectories with different initial conditions are depicted for $Q=0.3$. The initial conditions of blue, yellow and green orbits are set as $(X,Y,Z,W)=(0.001,\,-0.1,\,0.001,\,0.001),~(0,\,-0.1,\,0.01,\,0.02),~(0.001,\,-0.05,\,0.01,\,0.01)$, respectively. The blue point at $(X,Y,Z,W)=(0.00032,\,-0.04,\,0.054,\,0.00036)$ shows the fixed point corresponding to the anisotropic inflation, while the red point at $(X,Y,Z,W)=(0,\,-0.077,\,0, \,0.0013)$ shows the isotropic fixed point of the warm inflation.}
\label{Fig1}
\end{figure}

We plotted typical trajectories in FIG \ref{Fig1}.
Since the phase space consists of four dimensions, 
we chose three of four variables and plot three dimensional subspaces of 
$(X,Y,Z)$ (Left panel) and $(X,Y,W)$ (Right panel) for $Q=0.3$, $\kappa=50$, and $\lambda=0.1$.
We see the solution of isotropic warm inflation (Red dot) is a saddle point in these parameterization.
In fact, we found an unstable eigen state by a linear analysis around the isotropic fixed point. 
A trajectory first approaches the isotropic warm inflation but turns out to settle down to the fixed point of anisotropic warm inflation.

We considered power-law solutions where inflation never ends.  In the real universe, inflation ends after a finite time. Hence, depending on the strength of dissipation, the anisotropic inflation will not appear by the end of inflation. 
That is, the dissipation of inflaton field prevents the anisotropy from growing during inflation.

\begin{figure}[H]
\centering
 \includegraphics[width=16cm]{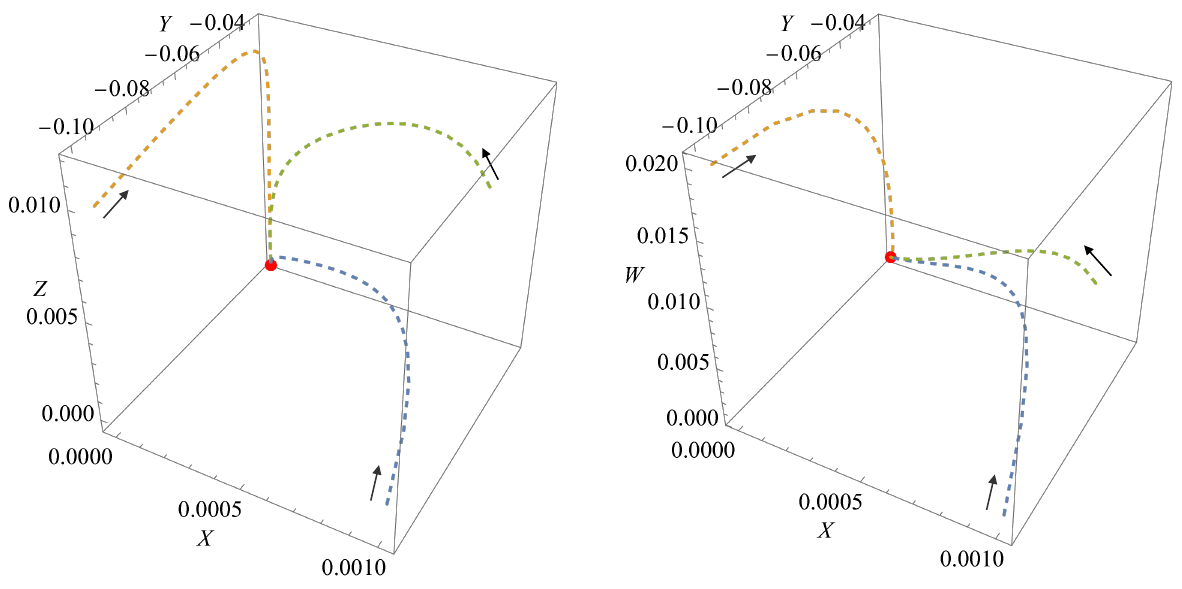}
 \renewcommand{\baselinestretch}{3}
 \caption{3D Phase space plots of attractors in the $(X,Y,Z)$-spaces (Left panel) and $(X,Y,W)$-spaces (Right panel). Trajectories with different initial conditions are depicted for $Q=3$. Initial conditions of each orbit indicated by the blue, yellow, and green dashed line are set as $(X,Y,Z,W) = (0.001,\,-0.1,\,0.001,\,0.001),~(0,\,-0.1,\,0.01, \,0.02),~(0.001,\,-0.05,\,0.01,\,0.01)$, respectively.
 The red point shows the isotropic fixed point at $(X,Y,Z,W)=(0,\,-0.025,\,0,\,0.0014)$ corresponding to the limit of the warm inflation.}
 \label{Fig2}
 \end{figure}

As $Q$ increses, the condition (\ref{Criterion}) gets hard to be satisfied.
We plotted the case of large $Q=3$ with $\kappa=50$, and $\lambda=0.1$ in FIG \ref{Fig2}. 
In this case, since the anisotropic fixed point (Blue dot) disappears, every orbit approaches the isotropic fixed point ($X=0$) that describes warm inflation. 
We also confirmed that
the fixed point is the attractor by using the linear analysis around the fixed point.
Thus, we found that the large dissipation tends to erase the anisotropy.

\section{Conclusion}%%%%%%%%%%%%%%%%%%%%%%%%%%%%%%%%%%%%%%%%%%%
In the era of precision cosmology, it is important to reconsider the cosmic no-hair conjecture \cite{Soda:2014awa,Maleknejad:2012as,Gao:2021qwl}. 
The cosmic no-hair conjecture supposes that cosmological spacetime obeys the Einstein equation with an inflaton field that behaves like a positive cosmological constant in the slow-roll limit. 
Curiously, if the inflaton field couples with other matter fields, the attractor of the conventional slow-roll inflation can be destabilized by the dissipation of the inflaton field or the matter fields coupled with the inflaton field. 
In warm inflation, the inflaton field produces the radiation energy density continuously during inflation, and in anisotropic inflation, the inflaton field sustains the gauge field. Thus it is expected that energy density of such matter fields survives during inflation against the cosmic no-hair conjecture. 
In this paper, we considered whether the cosmic no-hair conjecture still hold even if we considered the both effects of the dissipation of the inflaton field and the presence of gauge field coupled with the inflaton field. 

We first clarified the condition~(\ref{ani-condition}) for making anisotropy survived by using slow-roll approximation and found that the condition became tight in the presence of dissipation compared with the condition without the dissipation. 
Then, we investigated a solvable inflation model by choosing the inflaton potential in Eq.~(\ref{potential}) and the gauge kinetic function in Eq.~(\ref{function}). We found exact solutions which represent power-law warm inflation and power-law anisotropic warm inflation in Section.~\ref{section4}.
We also clarified the phase space structure of the  power-law inflation models in Section.~\ref{section5}. 
It turned out that whether the anisotropy during inflation survives or not depends on the the magnitude of $Q$. Let us fix the coupling constant $c$ in the gauge kinetic function $f(\phi)$. If the inequality $c>1+Q$ is satisfied, the anisotropic warm inflation is realized.
%for the model with $c>1+Q$. 
%If  $Q$ has a moderate value that satisfies $c \gtrsim 1+Q$, the anisotropy would be hard to grow enough in the duration of inflation. 
If the inequality $c\leq 1+Q$ is satisfied
, no anisotropic inflation occurs and isotropic warm inflation becomes attractor.
%speed of decaying process from inflaton field into matter field.  If the process is very slow, the anisotropic warm inflation is realized. If the process is slow, the anisotropy tends to be hard to grow enough in the duration of inflation. If the process is rapid, no anisotropic inflation occurs.

In this paper, we assumed the dissipation ratio $Q$ is constant. In general, the dissipation depends on the Hubble parameter, the inflaton field, temperature and the mass of the inflaton field. However, in the slow-roll limit, the dissipation ratio is almost constant during inflation. Hence, assumption that dissipation ratio is constant is legitimate to clarify the condition for the existence of a solution of anisotropic warm inflation.  To support the validity of the assumption, we showed qualitative feature does not change even for time dependent dissipation ratio in the Appendix.

%Let us focus on the Hubble parameter dependence. 
Since Q increases toward the end of inflation as the Hubble parameter decreases,  it would be more difficult to realize anisotropy near the end of inflation. Hence, it would be worth investigating more realistic models in detail. In particular, phenomenological consequences of the anisotropic warm inflation should be investigated in future work.

\section*{Acknowledgments}
S.\ K. was supported by the Japan Society for the Promotion of Science (JSPS) KAKENHI Grant Number JP22K03621.
J.\ S. was in part supported by JSPS KAKENHI Grant Numbers JP17H02894, JP17K18778, JP20H01902, JP22H01220.
K.\ U. was supported by the Japan Society for the Promotion of Science (JSPS) KAKENHI Grant Number 20J22946.

\section{Appendix}%%%%%%%%%%%%%%%%%%%%%%%%%%%%%%%%%%%
In this paper, for simplicity, we assumed that Q  is constant. Here, we assess how the time-dependence of $Q$ would affect the analysis in section 3.

In general, $Q=\Upsilon/3H$ slowly varies during inflation. We checked the case where $\Upsilon$ is constant numerically for chaotic inflation ($n=2$).
As shown in FIG 4, when Q is constant, the energy density of radiation grows firstly and, after a while, the anisotropy grows. As shown in FIG 5, when Q varies as the Hubble parameter decreases  toward the inflation end, the qualitative behavior is still similar to that of $Q=$ constant although quantitative differences exist. 
\begin{figure}[htpp]
\centering
 \includegraphics[width=16.5cm]{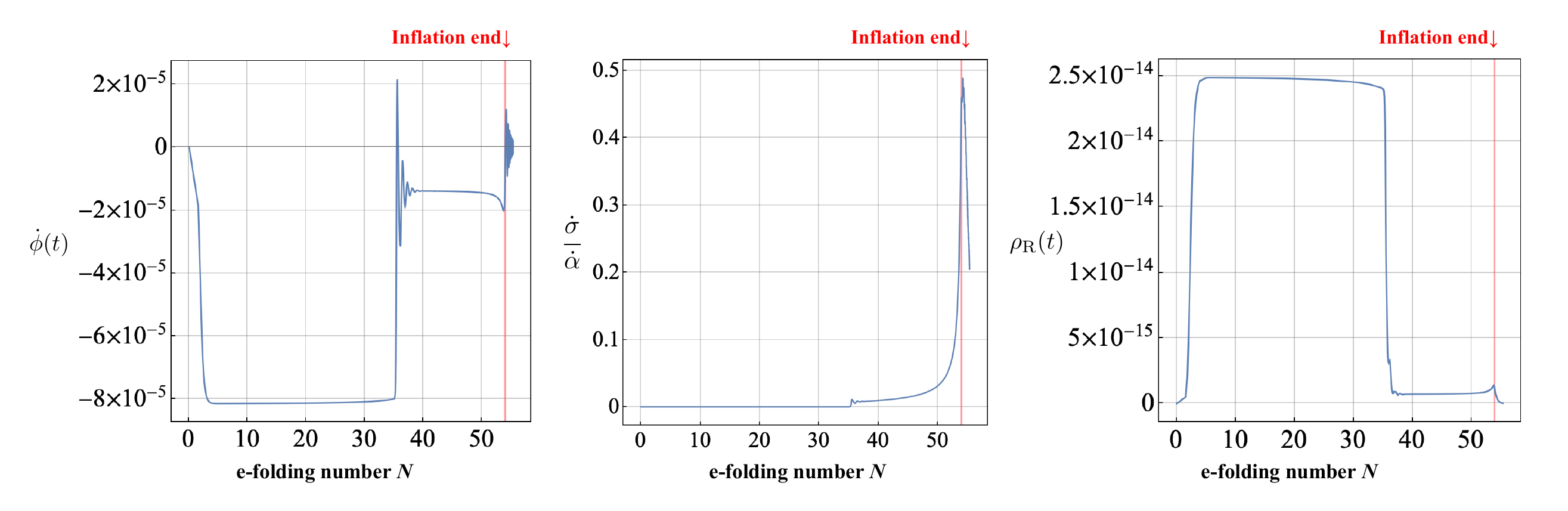}
 \caption{Assuming $Q$ is  constant, we calculated the velocity of the inflaton $\dot{\phi}$, the degree of anisotropy $\dot{\sigma}/\dot{\alpha}$, the energy density of radiation $\rho_R$ for chaotic inflation with $V(\phi)=\frac{1}{2}m^2 \phi^2$. The parameters are set as $Q=0.000005$ and $c=3$.}
\label{Fig4}
\end{figure}

\begin{figure}[htpp]
\centering
\includegraphics[width=16.5cm]{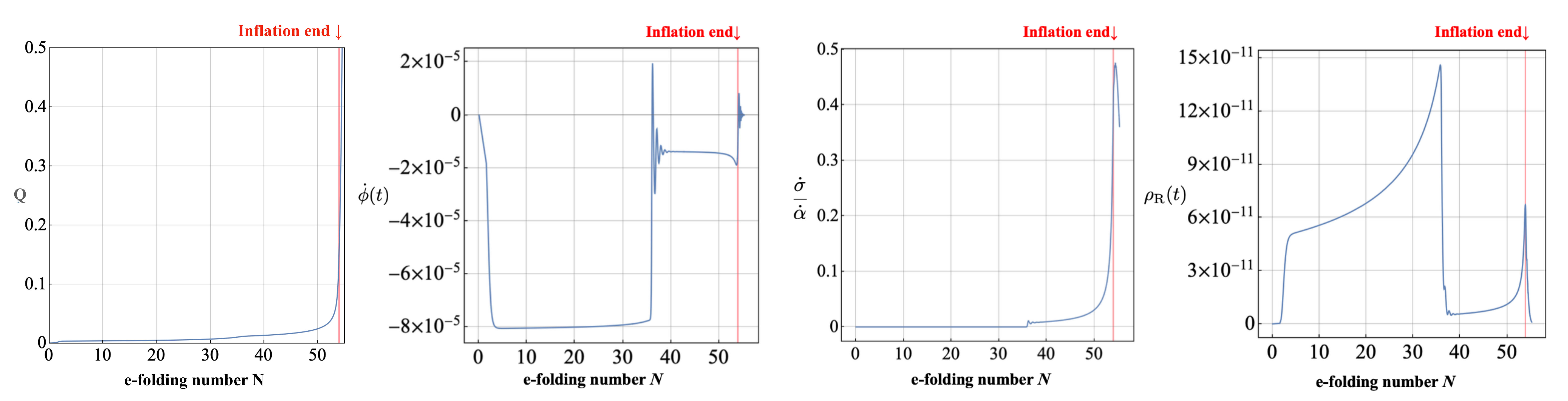}
 \caption{Assuming $\Upsilon$ is  constant, we calculated Q, the velocity of the inflaton $\dot{\phi}$, the degree of anisotropy $\dot{\sigma}/\dot{\alpha}$, the energy density of radiation $\rho_R$ for chaotic inflation with $V(\phi)=\frac{1}{2}m^2 \phi^2$. The parameters are set as $\Upsilon=0.000005 M_{\rm pl}$ and $c=3$.}
\label{Fig5}
\end{figure}
The point is that anisotropy will be realized even when Q varies, as long as the condition $c>1+Q$ is satisfied.
\printbibliography
\end{document}